\documentclass[a4paper, 12pt]{article}
\usepackage{graphicx}
\usepackage{amsmath}
\usepackage{amssymb}
\def \ni {\noindent}
\def \vs  {\vskip5mm}
\def \be {\begin{equation}}
\def \ee {\end{equation}}
\def \bea {\begin{eqnarray}}  
\def \eea {\end{eqnarray}}

\begin{document}    
\begin{titlepage}     
     
\title{Waiting for the Quantum Bus}     
     
\author{
A.J. Bracken
\footnote{{\em Email:} a.bracken@uq.edu.au} \, and
G.F. Melloy
\\School of of Mathematics and Physics\\    
The University of Queensland\\Brisbane, Australia 4072}
     
\date{}     
\maketitle     
\vs
\ni
%{\bf After $45$ years, an insight into a peculiar quantum effect sparks renewed interest.}     
\begin{abstract}
Forty-five years after the discovery of the peculiar quantum effect known as `probability backflow', and twenty years after
the greatest possible size of the effect was characterized, an experiment has been proposed recently to observe the effect in a Bose-Einstein condensate.  
Here the history is described in non-technical terms.

\end{abstract}

\ni
%PACS numbers:    
\end{titlepage}

%\section{Introduction}

Einstein's conviction that  ``God does not play at dice with the universe" is now generally seen as misguided.   
The quantum world, unlike our everyday classical world, is inherently probabilitistic in nature. 
But it may yet surprise the reader to learn that even in regard to probability itself,  
the quantum and classical worlds do not always behave in the same way.

Forty-five years ago, British physicist G R Allcock published three  papers \cite{allcock}, since widely-cited, on the so-called `arrival-time' 
problem in quantum mechanics. 
This is the problem of determining
the probability distribution over {\em times} at which a quantum particle may be observed at a given {\em position}.   
It  is converse to a problem that quantum mechanics deals with routinely, namely to predict the 
probability distribution over {\em positions}  at which a quantum particle may be observed at 
a given {\em time}.  The difference between the two looks so subtle as to be insignificant, but the  arrival-time problem is by no means as 
straightforward as its counterpart, and it   
has led to continuing argument and research 
over many years \cite{kijowski,aharonov,muga}. 

In his studies of the arrival-time  problem, Allcock noted  
a very peculiar quantum effect that has come to be known as `quantum probability backflow'. 
The effect has provoked continuing interest in its own right \cite{bracken1,bracken2,penz,yearsley,nielsen}, 
leading to a recent proposal to observe it directly by manipulating ultracold atoms  
in a Bose-Einstein condensate \cite{palmero}.

To appreciate the strangeness of quantum probability backflow, consider a comparable everyday phenomenon.  
Imagine a long straight road running past your front door, from far on your left to far on your right --- 
think of it as the $X$-axis,
with your house located at position $x=0$.   
You have been reliably informed that there is a 
bus somewhere on the road, possibly to your left, or possibly already to your right, 
but definitely travelling in a left-to-right direction. Moreover, the speed of the bus is definitely constant. 
 
Your informants were somewhat uncertain as to just what that speed is, 
but they were able to suggest a distribution of probabilities over possible left-to-right velocity values. 
Fig. 1 provides an example, where the velocity scale is miles per hour.   
In this example,   the bus is most likely travelling from left to right at about $20$ miles per hour.

%\vspace*{1in}

%Fig. 1 to go near here.

\begin{figure}[t]
\centering
\rotatebox{0}{\includegraphics[width=15cm]{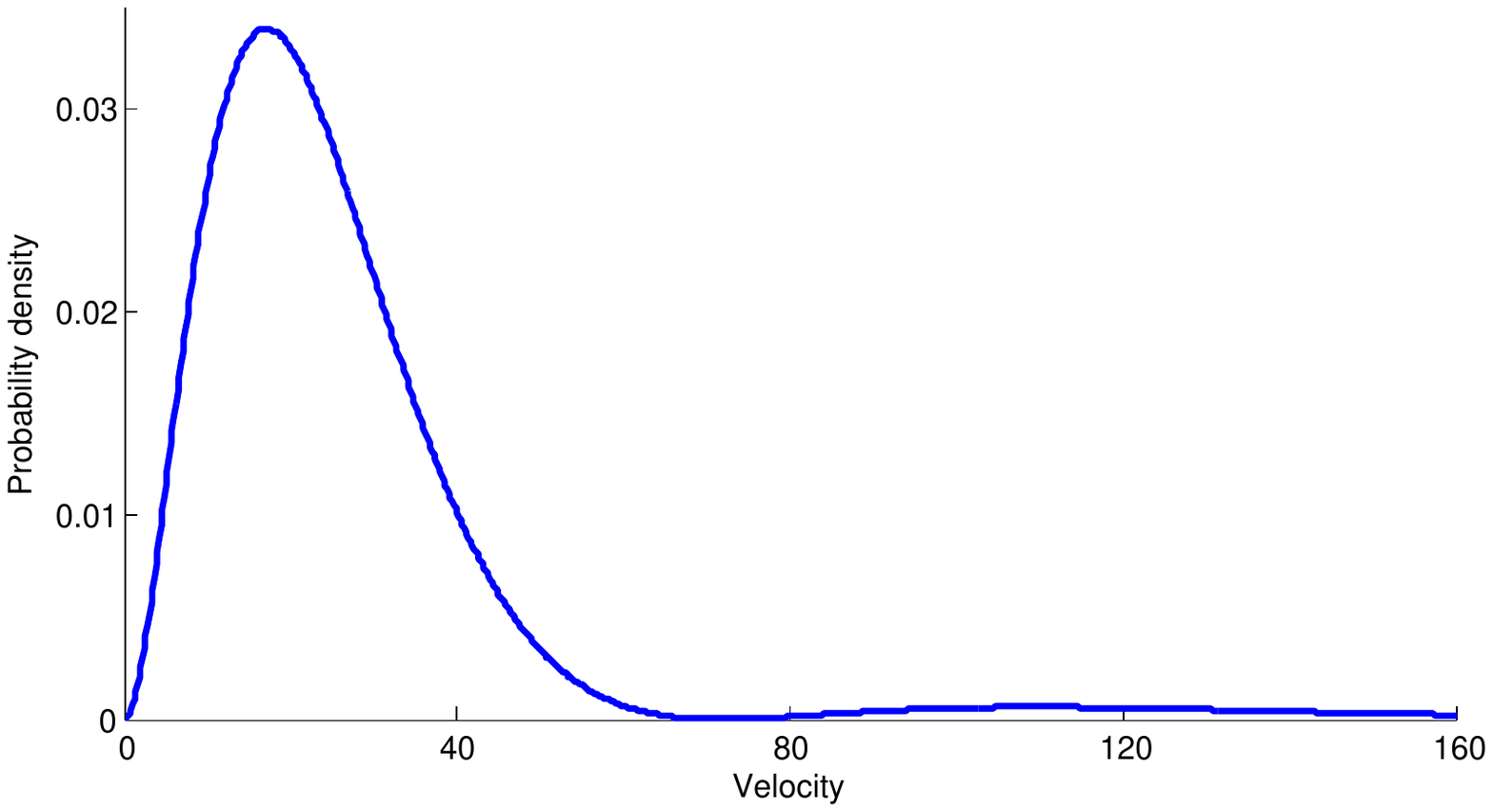}}
\caption{Distribution of probability over velocities in miles per hour/tens of microns per second of the classical/quantum bus.}
\end{figure}

%\vspace*{1in}

Your informants were also not quite sure  where the bus was located on the road at the time they spoke to you, but they were able to suggest 
another distribution of probabilities, over possible  positions of the bus along the $X$-axis at that initial time.
Fig. 2 shows an example, where the distance scale is miles. In this example there are matching probability distributions  to your left and right, and so a 
$50$-$50$ chance that the bus is still approaching at the initial time, and that it is about $2$ miles away.  

%\vspace*{1in}

%Fig. 2 to go near here.

\begin{figure}[t]
\centering
\rotatebox{0}{\includegraphics[width=15cm]{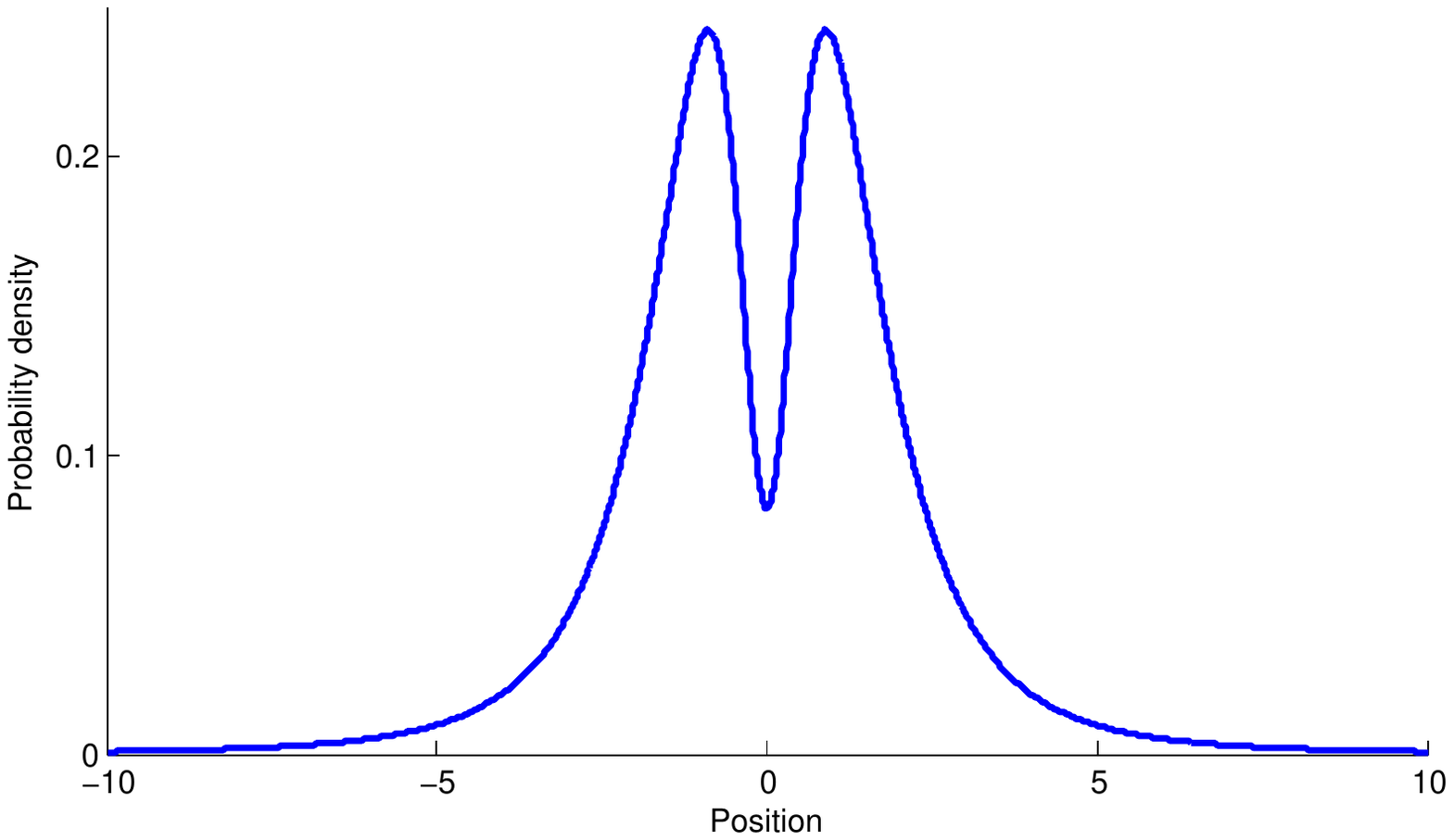}}
\caption{Distribution of probability over initial positions in miles/tens of microns of the classical/quantum bus.}
\end{figure}

%\vspace*{1in}

You wait by the road in the hope of flagging down the bus if and when it reaches you from the left.
The longer you wait, the more doubtful you become that the bus is still on its way.   
Common sense tells you that the 
probability of the bus being somewhere to your left is 
not increasing as time passes, given that it is travelling from left to right.   
 
Supporting your intuition, Fig. 3 shows distributions of probability over positions of the classical bus at successive times, 
starting from the assumed initial 
distribution as in Fig. 2, and assuming a distribution of velocities as in Fig. 1. 
The probability flows steadily towards the right, in the same direction as the velocity of the bus.  This is confirmed in Fig. 4, which shows the 
probability that 
the bus is still on its way from 
your left, starting from the value  $0.5$, or $50\%$, and decreasing steadily towards zero.     
  
%\vspace*{1in}

%Figs. 3 and 4 to go near here.

\begin{figure}[t]
\centering
\rotatebox{0}{\includegraphics[width=15cm]{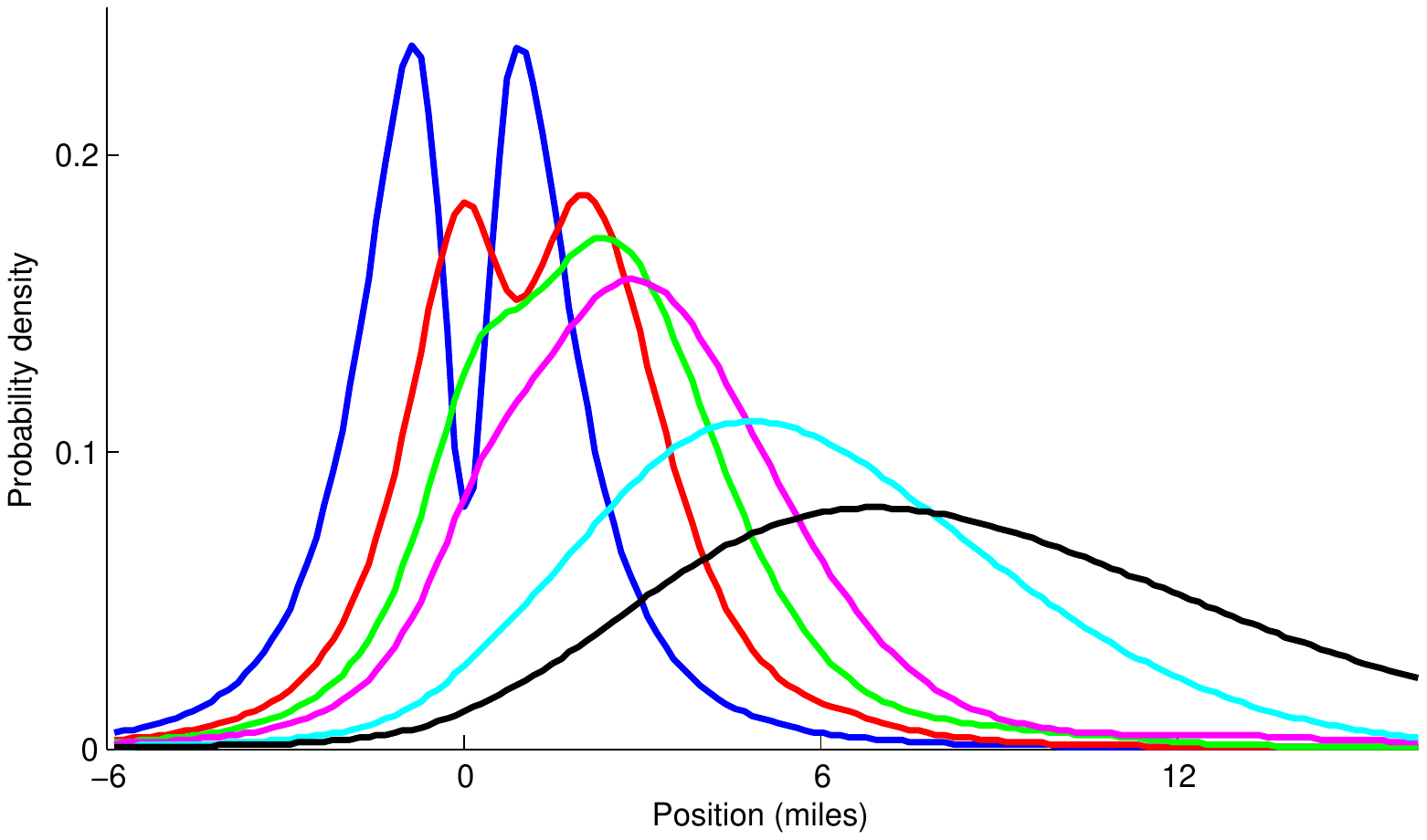}}
\caption{Distributions of probability over positions of the classical bus. 
From left to right, at times $0$ (blue), $0.05$ (red), $0.0875$ (green), $0.125$ (magenta), $0.25$ (cyan) and $0.375$ (black) hours.}
\end{figure}

\begin{figure}[t]
\centering
\rotatebox{0}{\includegraphics[width=15cm]{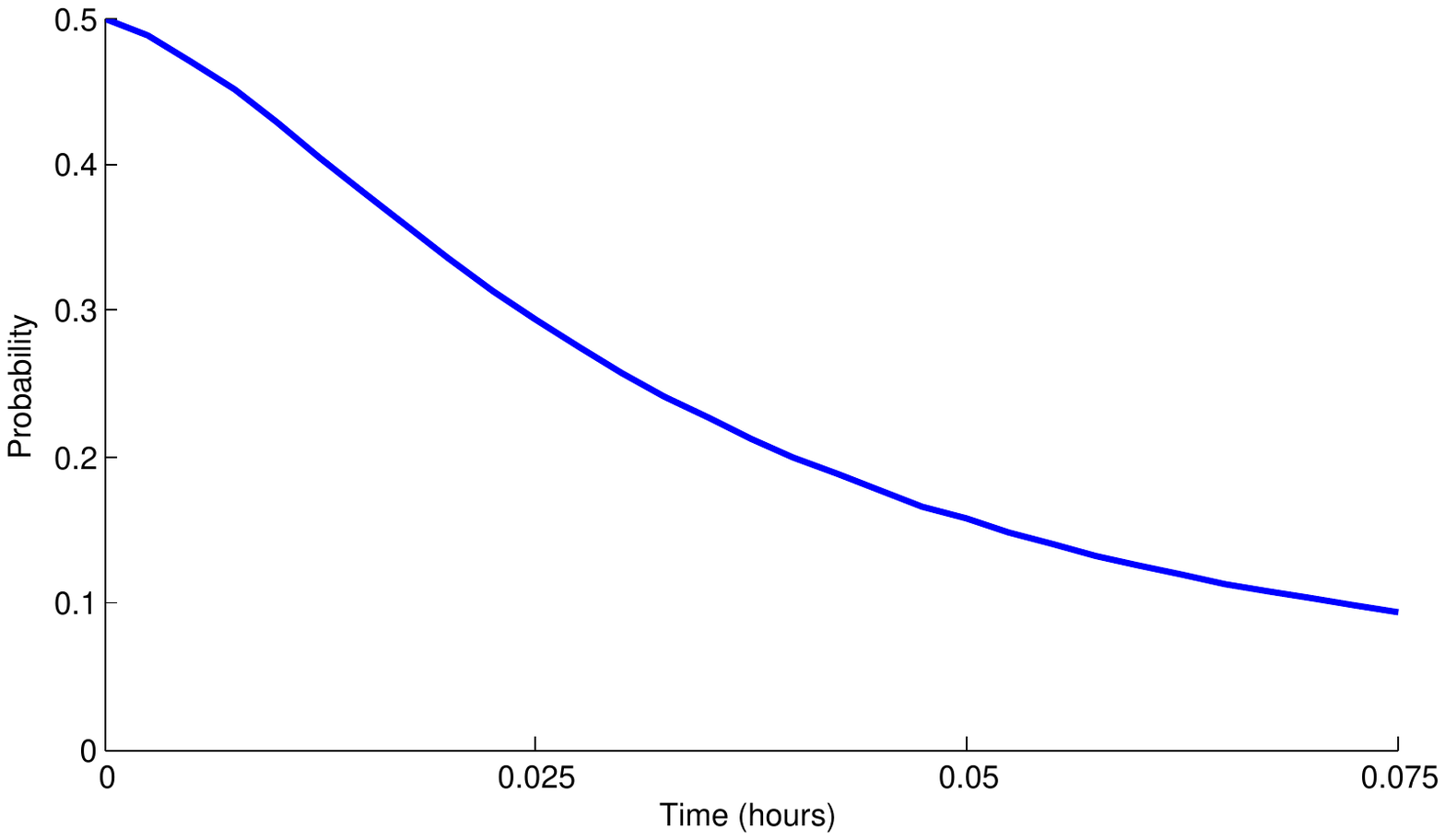}}
\caption{Probability that the classical bus is to the left of $x=0$, versus time in hours.}
\end{figure}

%\vspace*{1in}
 
Waiting for a `quantum bus' is different. To this end, consider  a free particle --- with a mass near that of an oxygen atom, say --- 
travelling   along the $X$-axis, and subject to the laws of non-relativistic quantum mechanics.  Such a quantum bus has a 
wave function satisfying Schr\"odinger's equation. This wave function can be chosen so that the distributions of 
probabilities over possible velocities and initial positions are again as in Figs. 1 and 2, 
where now distances are measured in tens of  microns, and 
velocities in tens of  microns per second.
  
More generally, all wave functions can be considered
for which  the quantum bus, like the classical one, 
has a (constant) velocity that is definitely directed from left to right, with uncertain magnitude, 
and an initial position that is also uncertain.  

While these conditions seem quite analogous to those of the classical bus,    
something completely counter-intuitive can happen in the quantum case.  

{\em Despite the fact that the bus 
is definitely travelling from left to right, the  probability of finding it  to your left on measuring its position, may 
increase as the time of measurement increases.} Because the bus is certainly somewhere 
on the line at all times, 
the probability of finding that
its position lies to your right must  decrease accordingly during this same time interval.   In other words, 
probability must flow backwards across position $x=0$, in the opposite direction to the velocity of the bus, 
even though there is no apparent force acting  
to `push the probability
backwards'.   
   
Fig. 5  shows possible probability distributions over positions of the quantum bus at successive times, 
arising from a choice of initial wave function  consistent 
with the probability distributions shown in Figs. 1 and 2.  The curves again show  movement of the probability distribution
from left to right, and are not greatly dissimilar to those shown for the classical bus in Fig. 3.  What is more, the graph in Fig. 6, 
showing the changing value of the probability that the quantum bus is still to your 
left, does not appear much different from its classical counterpart in Fig. 4.  It is easy to overlook  in Fig. 6 the rise and then 
fall of the probability value 
at very early times, up to about $0.002$ seconds. 

%\vspace*{1in}

%Figs. 5 and 6 to go near here.

\begin{figure}[t]
\centering
\rotatebox{0}{\includegraphics[width=15cm]{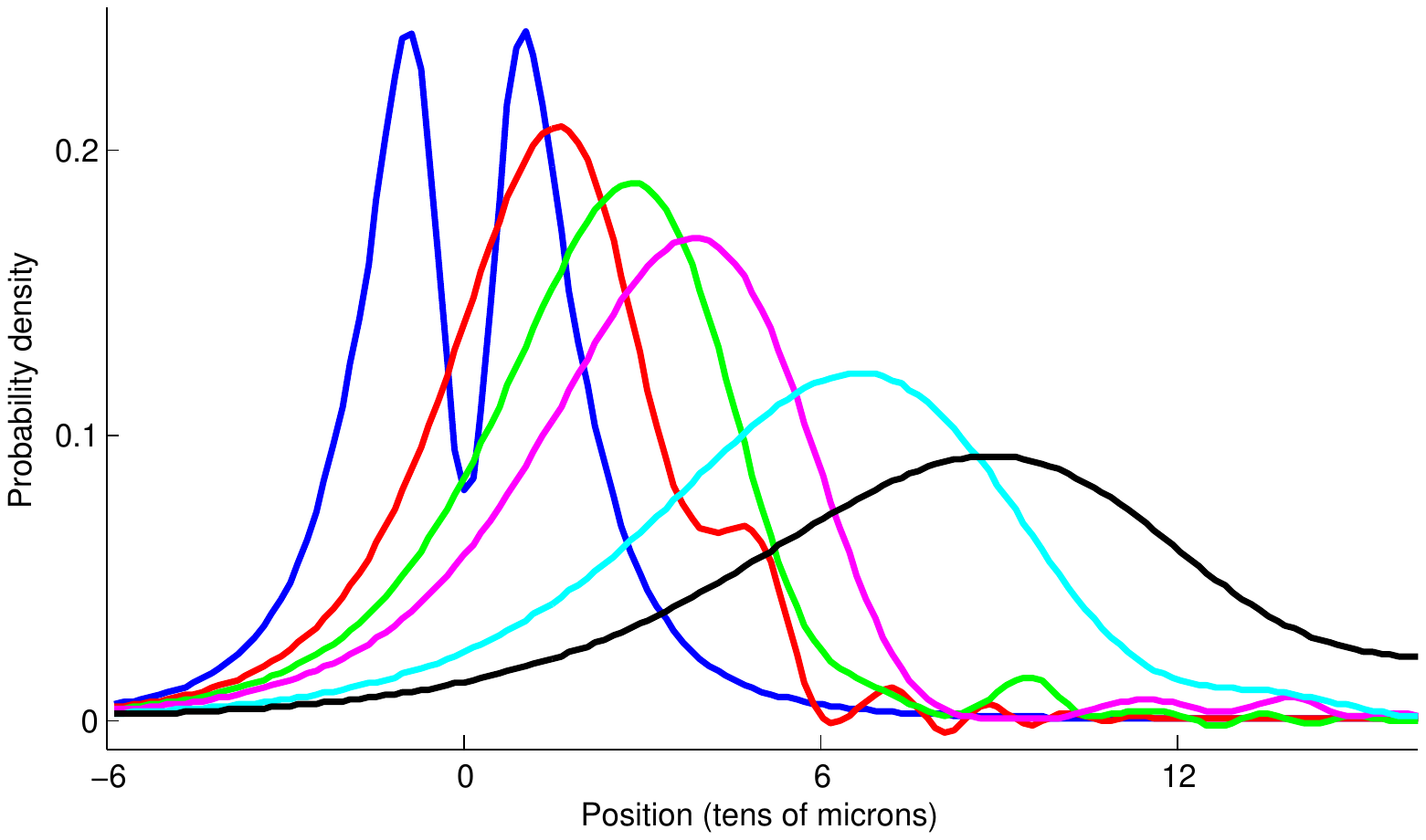}}
\caption{Distribution of probability over positions of the quantum bus.  
From left to right, at times $0$ (blue), $0.05$ (red), $0.0875$ (green), $0.125$ (magenta), $0.25$ (cyan) and $0.375$ (black) seconds.}
\end{figure}

\begin{figure}[t]
\centering
\rotatebox{0}{\includegraphics[width=15cm]{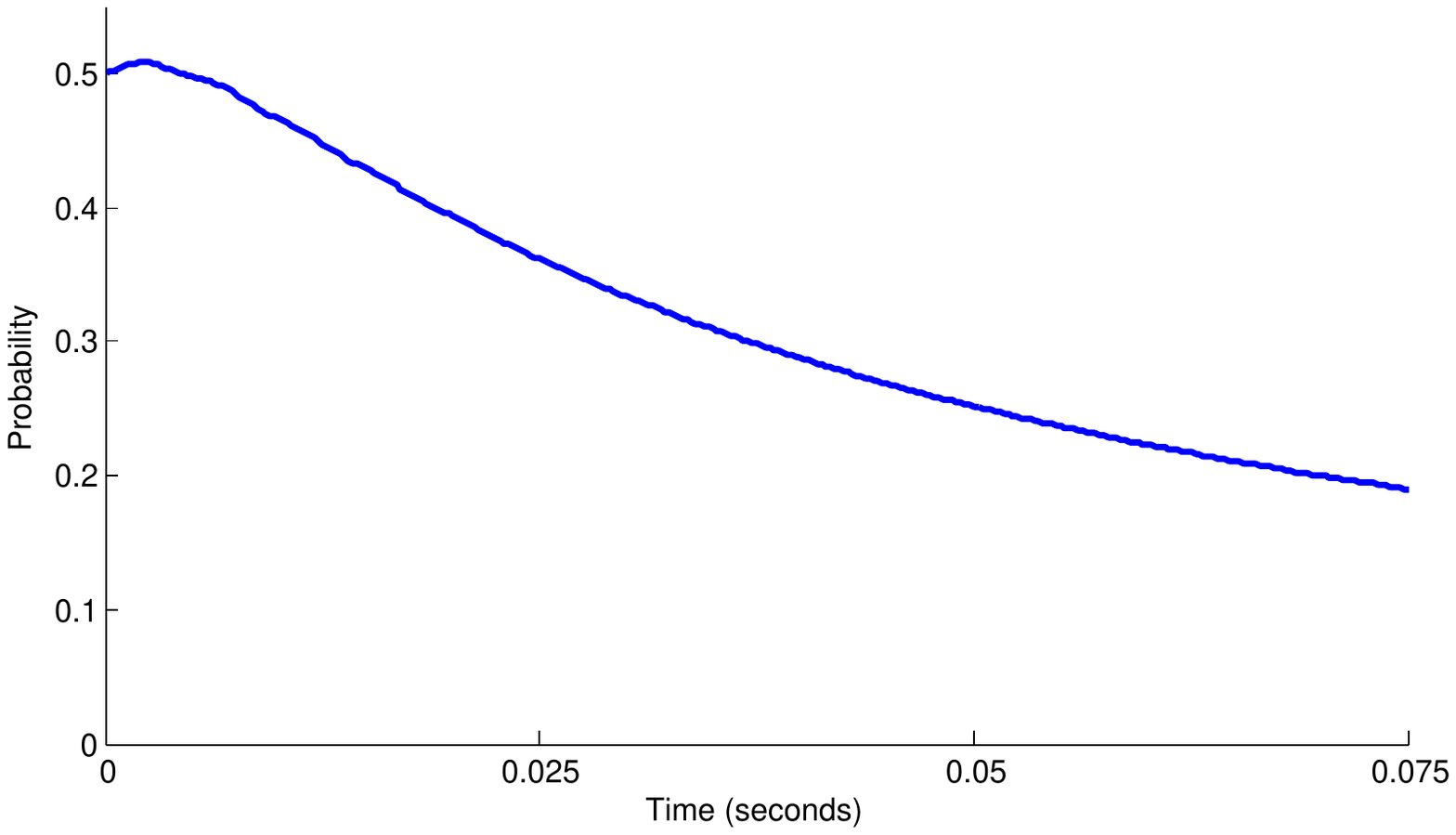}}
\caption{Probability that the position of the quantum bus is to the left of $x=0$, versus time in seconds.}
\end{figure}
  
%\vspace*{1in}

If we look more closely at what happens at these early times, we can  see the striking difference between the classical and quantum cases.
Fig. 7 shows plots of the probability distribution over positions of the classical bus, at times $0$ (in blue) and $0.002$ hours (in red).
The probability flows from left to right over this time interval, just as it does at later times, and this is confirmed in Fig. 8, 
which shows the changing value of the probability that the classical bus is still to your left.  Again, it decreases steadily.

%\vspace*{1in}

%Figs. 7 and 8 to go near here.

\begin{figure}[t]
\centering
\rotatebox{0}{\includegraphics[width=15cm]{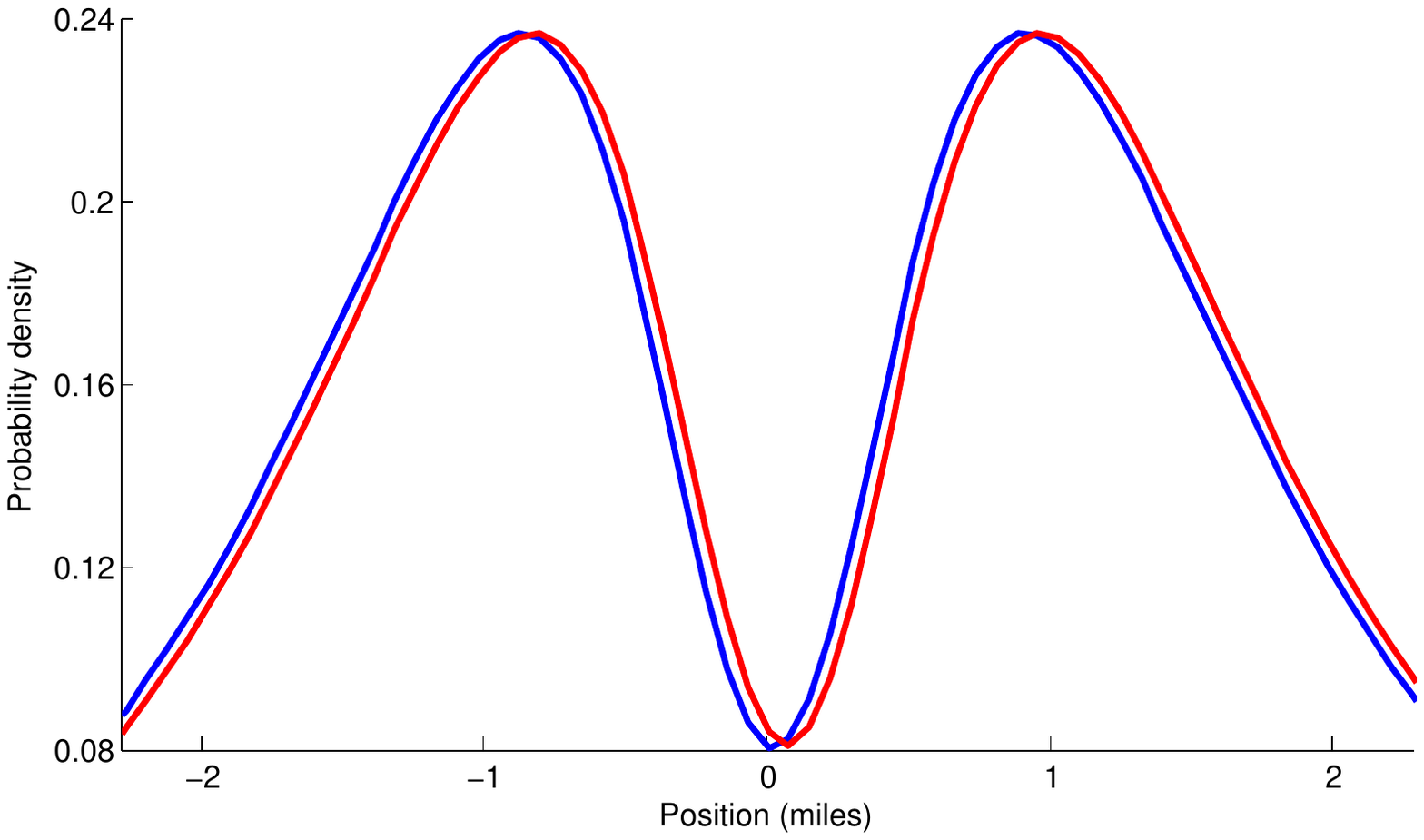}}
\caption{Distributions of probability over positions of the classical bus at times $0$ (blue) and $0.002$ (red) hours.}
\end{figure}

\begin{figure}[t]
\centering
\rotatebox{0}{\includegraphics[width=15cm]{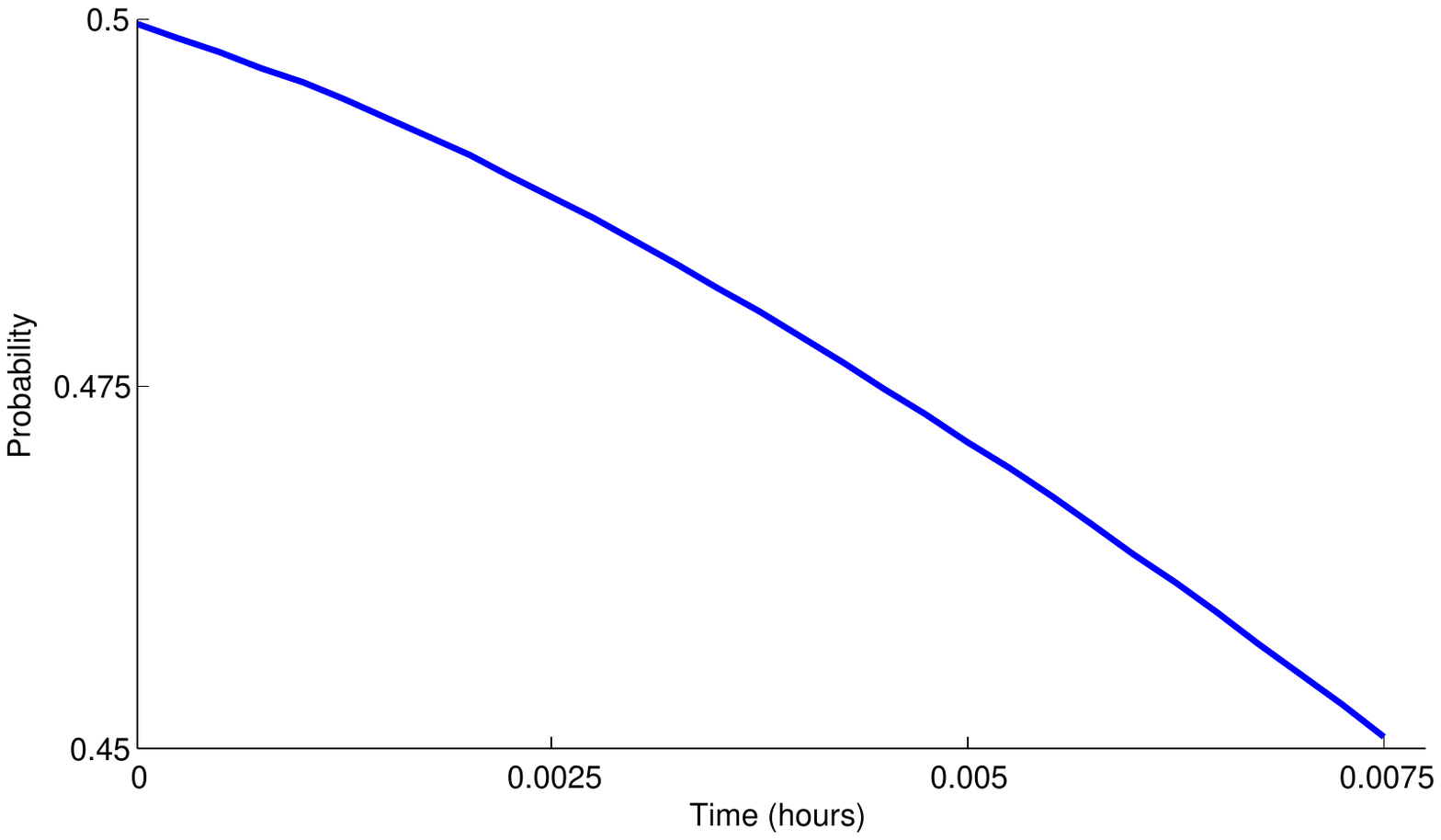}}
\caption{Probability that the classical bus is to the left of $x=0$, versus time in hours.}
\end{figure}
 
%\vspace*{1in}

Figs. 9 and 10 show the analogous results for the quantum bus.  While the successive curves in Fig. 9 are again {\em suggestive} 
of movement to the right during
the time interval of length $0.002$ seconds, Fig. 10 shows that 
the probability that the position of the bus will be found on your left, actually increases during this period, from  the initial value of 
$0.5$ to a value of about $0.508$, before starting to decline.  In other words, while there is a $50\%$ probability initially 
that the position of the bus will be found on your left, 
this has increased to about $50.8\%$ after about $0.002$ seconds.  

%\vspace*{1in}

%Figs. 9 and 10 to go near here.

\begin{figure}[t]
\centering
\rotatebox{0}{\includegraphics[width=15cm]{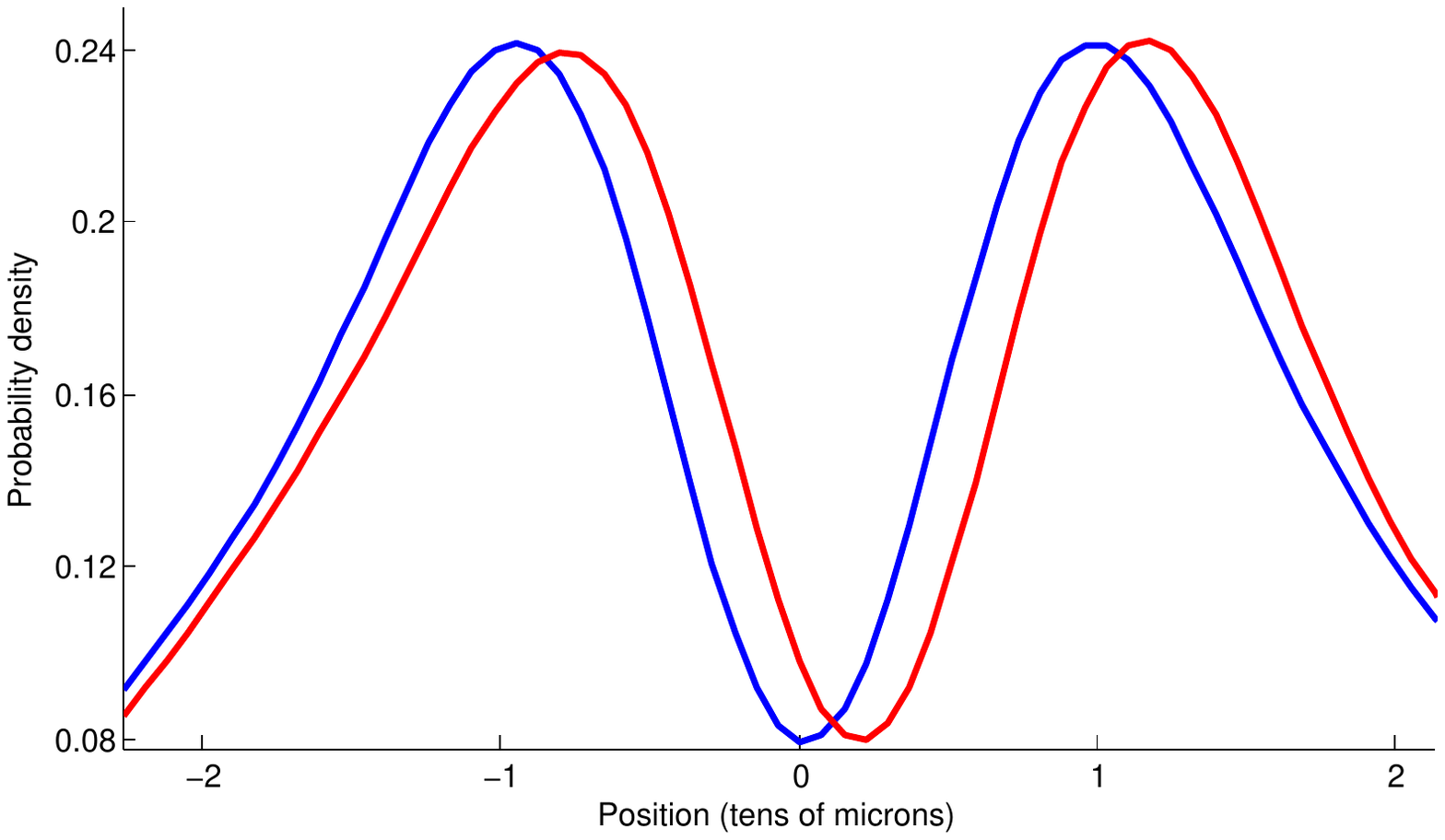}}
\caption{Distribution of probability over positions of the quantum bus at times $0$  (blue) and $0.002$ (red) seconds.}
\end{figure}

\begin{figure}[t]
\centering
\rotatebox{0}{\includegraphics[width=15cm]{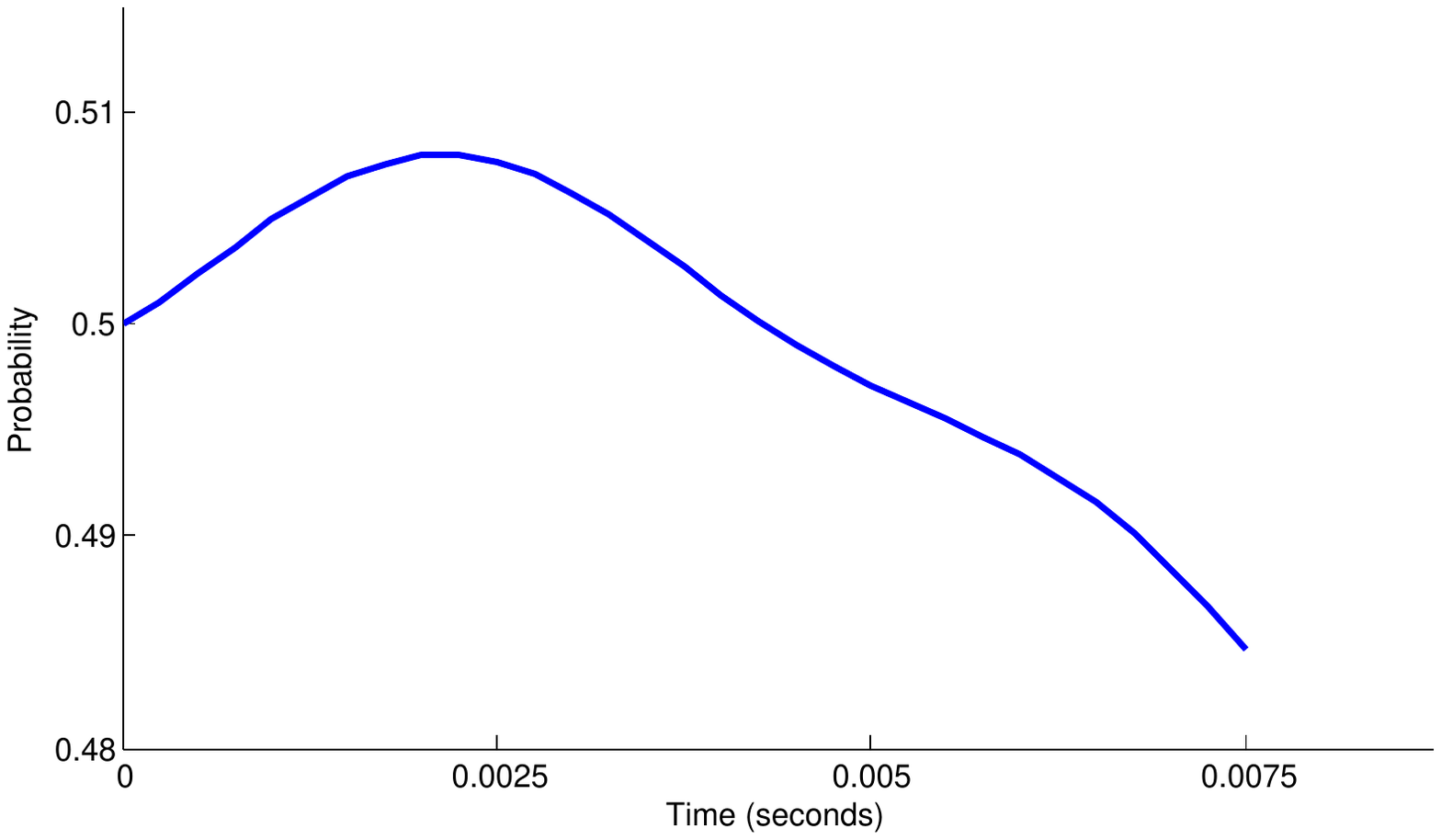}}
\caption{Probability that the quantum bus is to the left of $x=0$, versus time in seconds.}
\end{figure}

%\vspace*{1in}

For each curve like those in Figs. 5 and 9, the area under the curve to the left of position $0$ gives the probability that the position 
of the quantum bus will be found on your left
at the corresponding time, while the area under the curve to the right gives the probability that it will be found 
on your right.  And for each curve, the sum
of these two areas equals one. 
  
For the blue curve in Fig. 9, these areas are each equal to $0.5$.  For 
the red  curve, however,  the area to the left equals about $0.508$, and so is greater than the area to the right, which equals about $0.492$.  This is
despite the appearance given by the red curve of 
motion to the right from the blue curve.  The apparent motion to the right is more than compensated 
for by a broadening of 
the left hand peak, and a narrowing of the right hand one.

From Allcock's work it is clear  
that probability backflow is 
an interference effect, made possible because a  
quantum particle, unlike a classical bus,  also has wave-like properties.   The appearance of mathematically similar `retro-propagation' effects is not 
uncommon in classical wave theories \cite{berry}.  The surprise in the quantum context comes not so much from unusual mathematics, as
from the fact that it is probability
that flows backwards in this case, counter to our classical intuition. 
 
The probability distributions over initial positions and  velocities can be chosen independently for a classical bus, 
whereas for a quantum particle, they are 
tightly related, being determined from  the wave function and its Fourier transform.  This relationship involves Planck's constant $h$ and,
as time evolves, also the
mass of the particle, and it   
is this same relationship between wave function and Fourier transform that
gives rise to that most profound of all quantum effects, Heisenberg's uncertainty principle. 

It is not hard to appreciate the relevance of probability backflow to the arrival-time problem that was 
primarily of interest to Allcock forty-five years ago.  
For if the probability
of finding a right-moving particle to the left of a given point can increase or decrease with time, the very notion of the probable 
time of its arrival 
at that point, from the left say,  may become problematic. 

In an attempt to raise interest in the backflow effect itself, and to quantify it, two aspects were considered by us some twenty-five years after
Allcock's work \cite{bracken1}.
The first aspect concerns the possible duration of the effect.  It is clear enough, and easily proved, that the probability of finding the
position of the 
quantum bus to the left of a given point such as $x=0$, must
approach zero for sufficiently long times, while the probability of finding it to the right of that point must approach one, 
just as in the case of the classical bus.     
We are then led to ask,  for how long can the quantum 
probability flow backwards past a given point? 

The answer to this question is simple, if surprising.   Backflow can occur over any finite time-interval, no matter how long.  
For once a quantum state is found that produces the effect over a time-interval of given duration,  another state can be constructed 
that produces the same amount of probability backflow over an interval twice as long, say, by a simple scaling process.  

The second and   more interesting question is  to ask: What is the greatest amount of probability --- call this amount $P$, say ---
that can flow `backwards' over a given time-interval?  The length of the time interval is immaterial, by the scaling argument just mentioned.  
This second question leads to a well-defined mathematical problem, but one  that has, to date, defied exact solution.   
Our estimate of $P$, obtained by treating the problem numerically,  was about $0.04$, or about $4\%$ of the total probability on the line.  
Subsequent numerical studies \cite{penz,yearsley} have found the more precise estimate $0.0384517...$. 
This is about $5$ times greater than the amount of 
backflow in the example treated above.    

The number $P$ appears as an eigenvalue in the mathematical problem that arises, and the corresponding eigenfunction defines
the wave function of the quantum state that gives rise to the greatest backflow value.  While eigenvalue problems commonly arise 
for observable quantities in 
quantum mechanics, leading to quantum numbers such as the energy levels of an atom, for example,  the eigenvalue 
$P$ is quite unusual.  

It is a pure (dimensionless) number.  Furthermore, its value is not only independent of the length of the time-interval over 
which probability backflow occurs, as already noted, 
but is also  
independent of the mass of the bus and, more surprisingly, 
of the  value of Planck's constant $h$, which typically characterizes quantum effects.  In an imaginary world where 
$h$ had a value 
many times larger or many  times smaller than the one we are familiar with in our world, 
the maximum possible probability backflow for a free quantum particle would still be the same, namely $0.0384517...$\,.   

A striking interpretation of probability backflow can be given if we introduce negative probabilities 
into the quantum description.  While such a notion may seem absurd, the utility of this radical concept for the conceptualisation of other aspects of 
quantum mechanics has been argued long ago in other contexts by Dirac \cite{dirac} and Feynman \cite{feynman}.  
The idea of negative probabilities seems less absurd when we recall 
how useful a related notion has become in everyday life --- the notion of negative numbers.
No-one ever saw a negative amount of money, for example, 
but there is a simplifying conceptual advantage in regarding withdrawals from your bank account as negative deposits 
when calculating the balance remaining
after all transactions are considered.  
Provided that your final balance is not negative, your bank manager will not complain that you used 
unphysical negative numbers to help you keep track.     

To see the relevance of negative probabilities in the present context, reconsider firstly the classical bus.  
Rather than distinct probability distributions over initial positions and velocities, 
we can consider a single distribution over both.  For the classical state illustrated in Figs 1 and 2, a 3D plot of 
such a joint 
distribution could look initially as in Fig. 11. 

%\vspace*{2in}

%Fig. 11, 12 and 13 to go near here.

\begin{figure}[t]
\centering
\rotatebox{0}{\includegraphics[width=15cm]{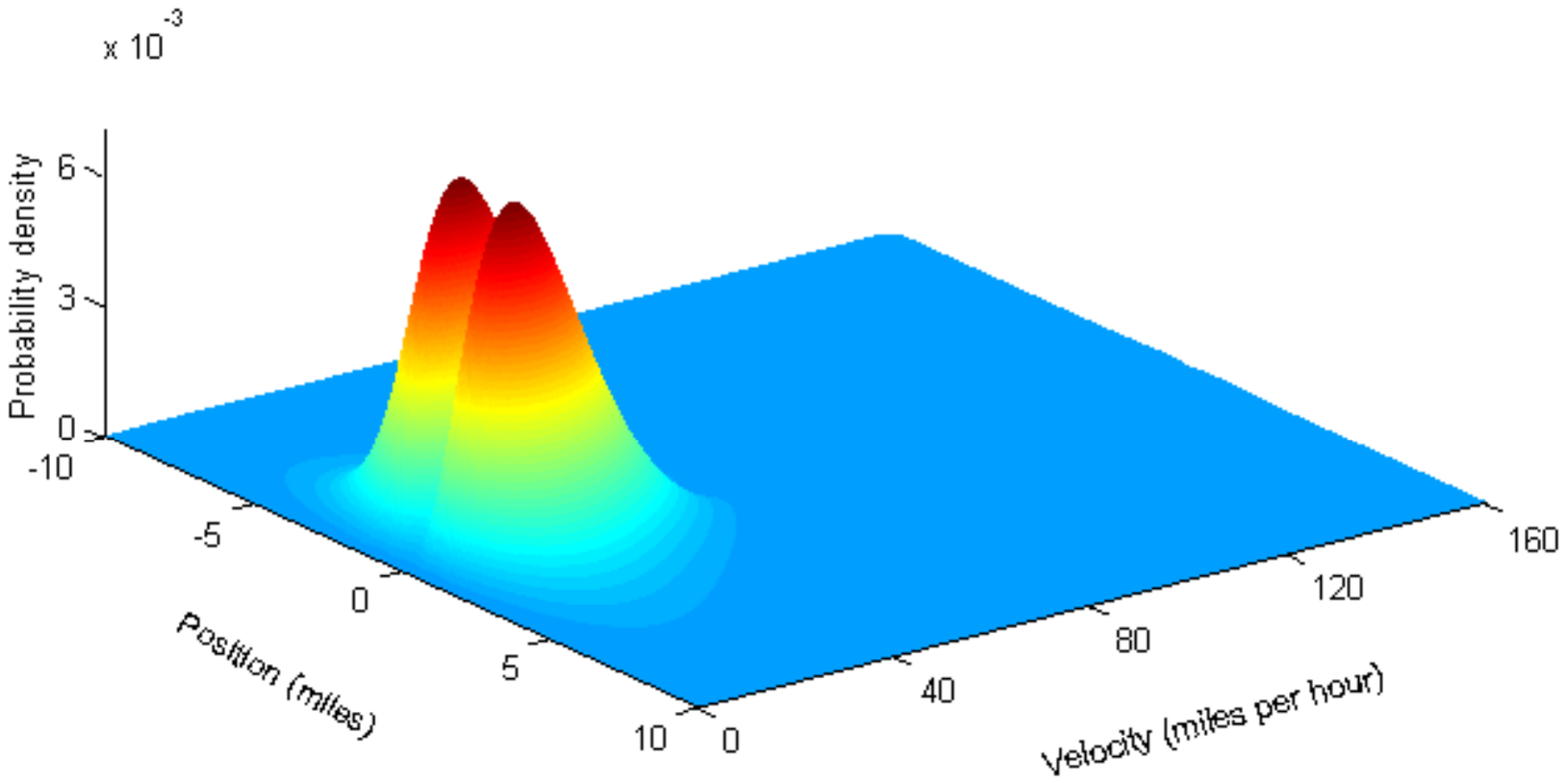}}
\caption{Distribution of probability over velocities and initial positions of the classical bus: 3D plot.}
\end{figure}

%\vspace*{1in}

\begin{figure}[t]
\centering
\rotatebox{0}{\includegraphics[width=15cm]{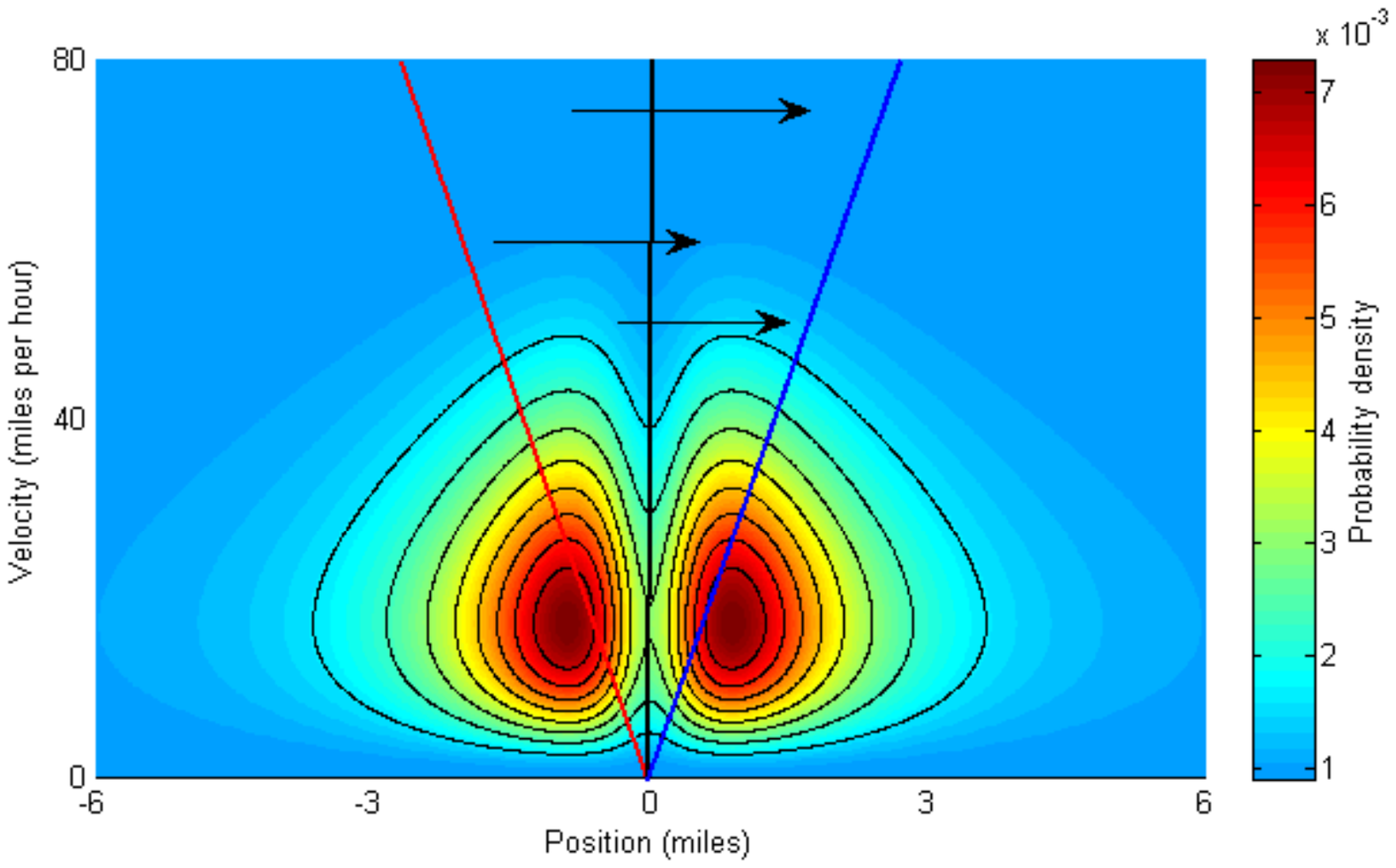}}
\caption{Distribution of probability over velocities and initial positions of the classical bus: contour plot.}
\end{figure}

%\vspace*{1in}

\begin{figure}[t]
\centering
\rotatebox{0}{\includegraphics[width=15cm]{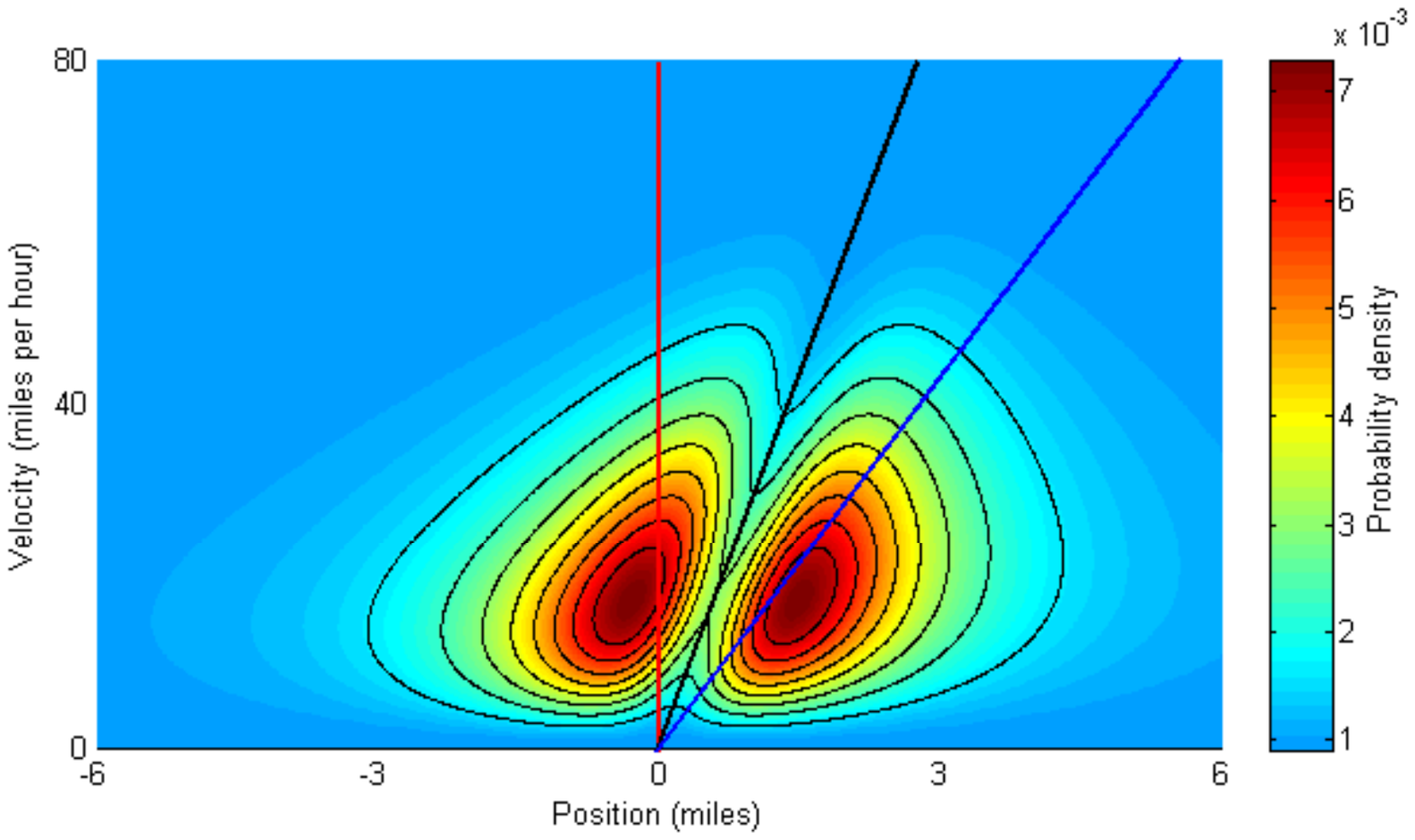}}
\caption{Distribution of probability over velocities and positions of the classical bus at a later time: contour plot.}
\end{figure}

Note that the distribution is confined to the upper half of the $XV$-plane because the bus 
definitely has a positive velocity.    Fig. 12 shows a contour plot of the initial distribution. 
The probability that the bus is initially to the right of $x=0$ is 
now given by the total probability 
distributed on the right quadrant of Fig.12, while the probability that the bus is to the left of $x=0$, and so yet to 
arrive at $x=0$, is given by the total probability distributed on the left quadrant.   The sum of these two probabilities equals $1$.

As time passes, the distribution on the upper half of the $XV$- plane evolves by shearing to the right, as shown in Fig. 12, leading after some time  
to Fig 13.  
All the probability in the wedge shown in Fig. 12, bounded by the red line and the 
black line, flows across the black line $x=0$ during this time interval, replacing the probability 
in the wedge bounded by the black line and blue line, which moves further right, as shown in Fig. 13.  
In this way the probability in the right quadrant 
increases while 
the probability in the left quadrant decreases by the same amount.  
 
For the quantum bus, the closest analogue of a joint distribution of position and velocity probabilities is given by a construct from the 
wave function due to Wigner \cite{wigner}.  For the quantum state giving rise to Figs. 1 and 4, 
the 3D  plot of this Wigner function, so-called,  might look initially as shown in Fig. 14.

Figs. 14, 15 and 16 to go near here.

%\vspace*{2in}

\begin{figure}[t]
\centering
\rotatebox{0}{\includegraphics[width=15cm]{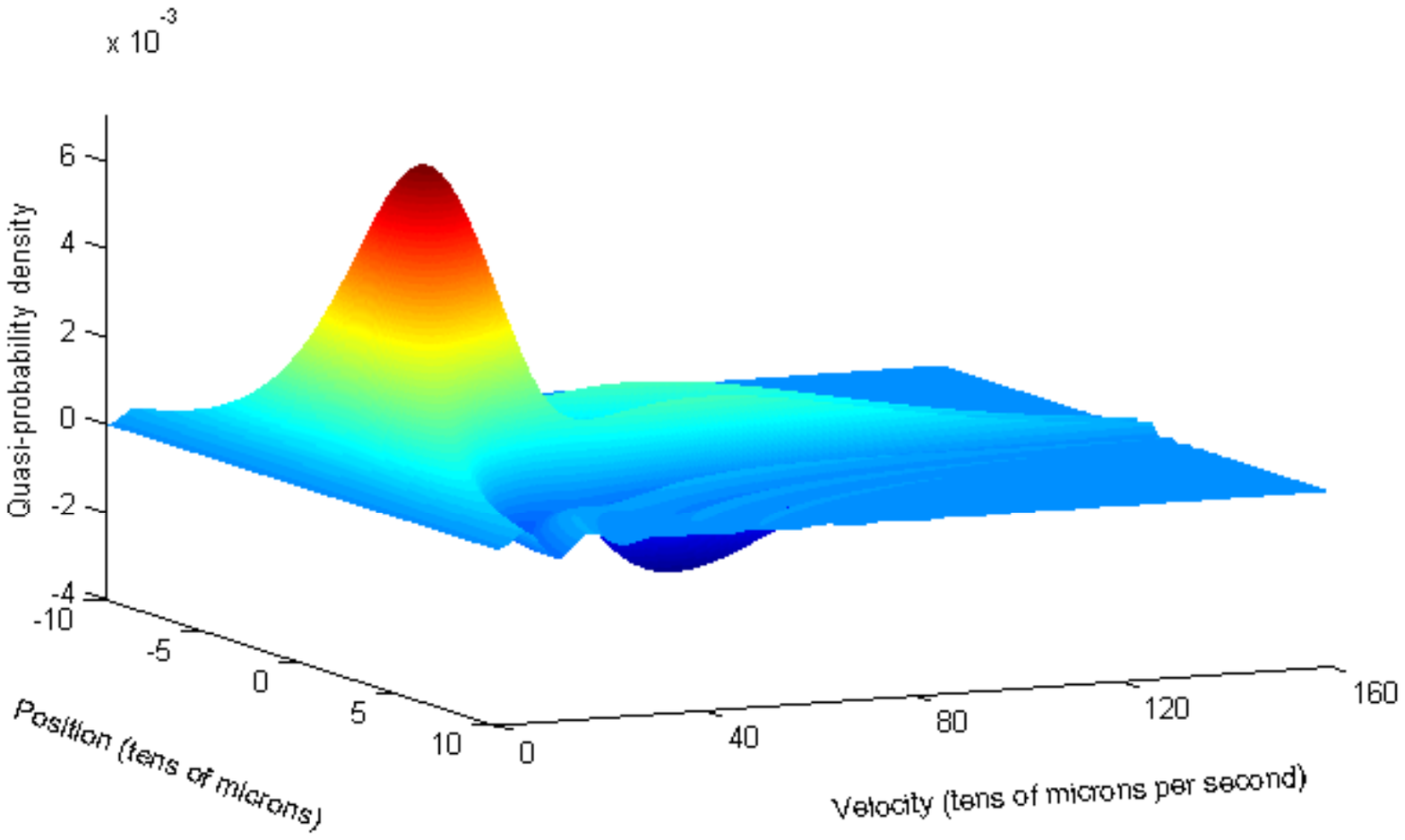}}
\caption{Distribution of quasi-probability over velocities and initial positions of the quantum bus: 3D plot.}
\end{figure}

%\vspace*{1in}

\begin{figure}[t]
\centering
\rotatebox{0}{\includegraphics[width=15cm]{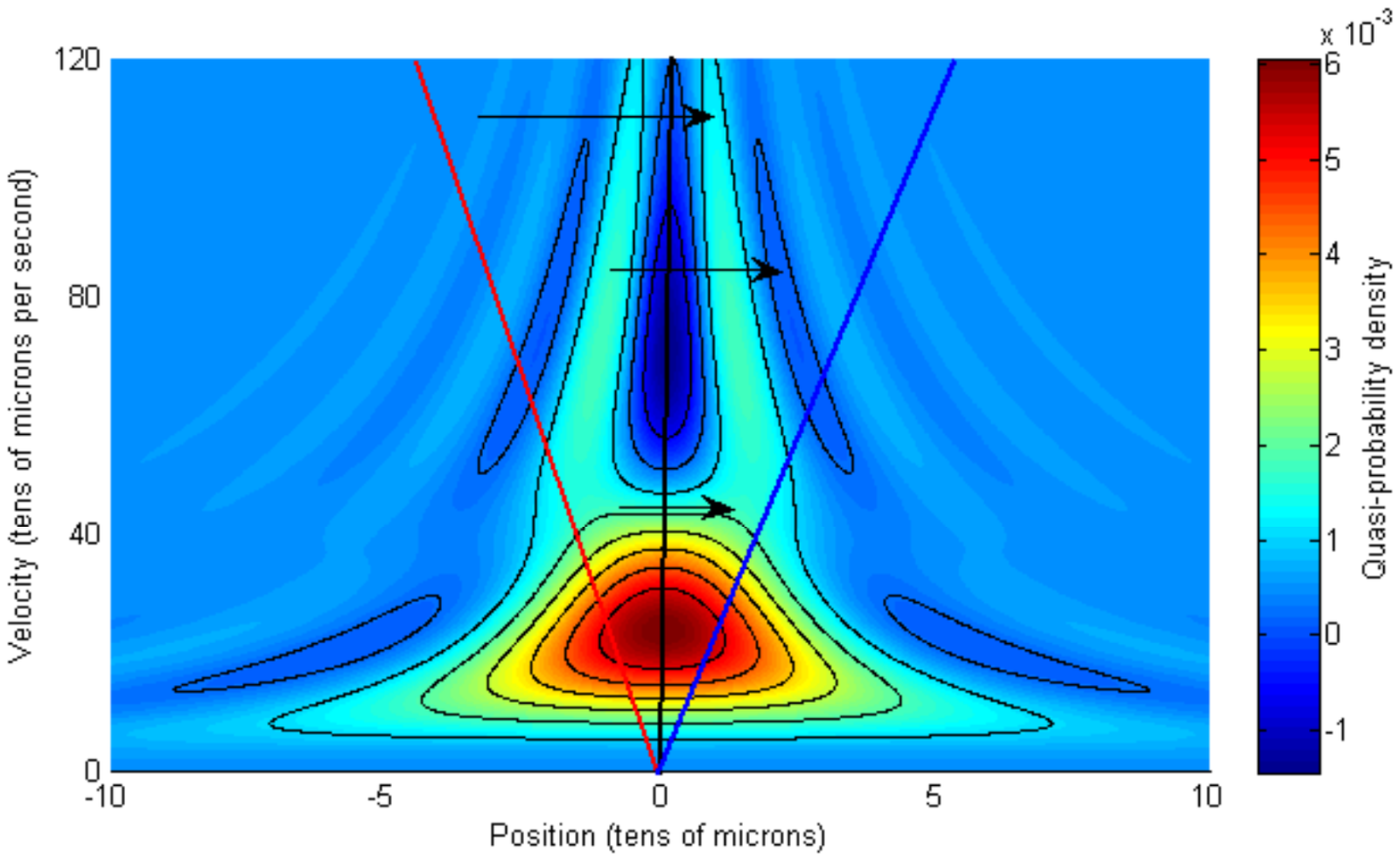}}
\caption{Distribution of quasi-probability over velocities and initial positions of the quantum bus: contour plot.}
\end{figure}

%\vspace*{1in}

\begin{figure}[t]
\centering
\rotatebox{0}{\includegraphics[width=15cm]{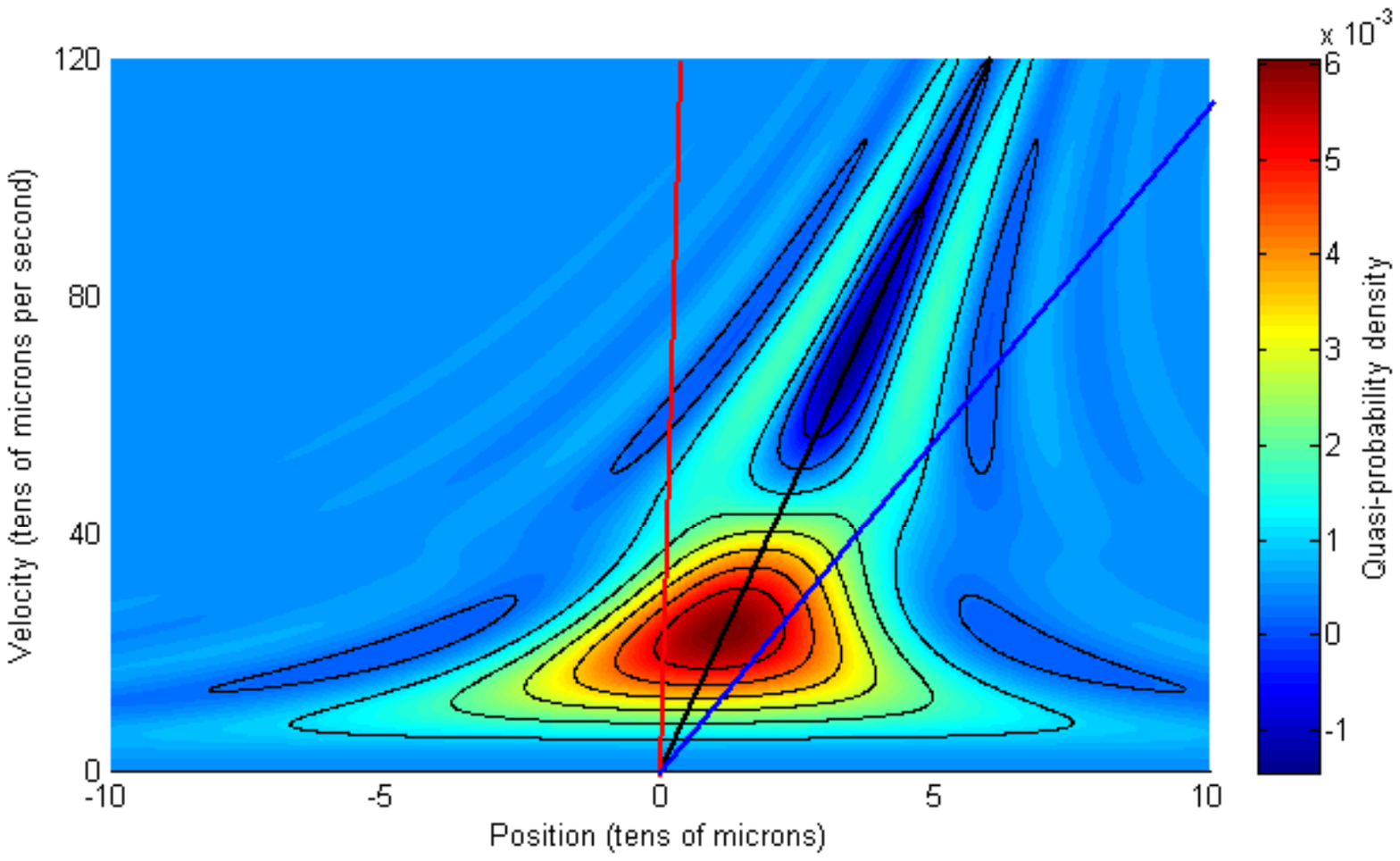}}
\caption{Distribution of quasi-probability over velocities and positions of the quantum bus after $0.05$ seconds: contour plot.}
\end{figure}

The Wigner function is not everywhere positive, unlike its classical counterpart in Fig. 11.  
Negative values appear in dark blue in the example shown in Fig. 14.  For the same example, 
Figs. 15 and 16 show contour plots of the initial distribution
and the distribution at a later time, with regions of negative values again shown in dark blue.     

Despite the appearance of negative values,  the Wigner function  does share important 
properties with the classical distribution.   Thus, in the present context, it vanishes for negative velocities.  
Furthermore, the total of its values in the right 
quadrant equals the probability that the bus is to the right of $x=0$, the total of its values in the left quadrant equals the probability 
that the particle is to the left of $x=0$, and the sum of these two equals $1$.  We may say that the Wigner function defines a 
quasi-probability distribution, taking both positive and negative probability values. 
   
What is most important for the discussion of probability backflow is that, as time passes, the Wigner function evolves in just the 
same way as the classical distribution, by shearing to the right, as indicated by Figs. 15 and 16.  
In the quantum case, however, the total amount of probability on a wedge like that between the red and black lines in Fig. 15 can be negative.  
When this negative probability flows to the right quadrant, arriving between the black and blue lines after some time, 
the probability on the right quadrant quadrant decreases, and the 
probability on the left quadrant increases by the same amount.   

We may say that all the probability moves to the right with the
motion of the quantum bus, just as in the classical case, but now not all that probability is positive.  Negative probability moving to the right has the 
same effect on the total probabilities in the left and right quadrants as positive probability moving to the left, 
thus giving rise to the backflow phenomenon.   Note that the greatest possible  
negative value of the total probability on any wedge like the one on the left in Fig. 15, evaluated over all quantum states 
with positive left-to-right velocities, equals the number $-P$.  
For the example given here, it has the value $-0.008...$, only about $20\%$ of    the maximum possible.

The value of $P$ evidently reflects the structure of the quantum description of a free particle, in terms of a 
wave function satisfying Schr\"odinger's equation, rather than the values of the only constants appearing in that description,
namely the mass of the particle and Planck's constant. 
Experimental observation of the value of $P$ would therefore provide some novel verification of this general mathematical framework.       
   
Successful observation of probability backflow in experiments, such as that proposed by  Palmero {\em et al.}, 
would  confirm Allcock's insight from forty-five years ago.  But a    
further challenge to 
experimenters would remain ---  
to measure the  mysterious quantity $P$ and verify that it has the value $0.0384517...$ predicted by quantum mechanics.

\vs

\end{document}